\documentclass[
 reprint,
 amsmath,
 amssymb,
 aps,
 prl,
 noeprint
]{revtex4-2}
\usepackage{hyperref}
\usepackage{siunitx}
\usepackage{graphicx}
\usepackage{subfigure}
\usepackage{bm}
\usepackage{xcolor}
\usepackage{soul}
\usepackage{titlesec}
\usepackage{listings}

\usepackage{dcolumn}
\usepackage{multirow}
\hypersetup{
    colorlinks=true,
    linkcolor=blue,
    citecolor = blue,
    filecolor=blue,
    urlcolor =blue
}

\definecolor{backcolour}{rgb}{0.95,0.95,0.92}
\definecolor{codegreen}{rgb}{0,0.6,0}

\DeclareMathOperator{\Tr}{tr}
\newcommand{\Rb}{\textsuperscript{87}Rb}

\newcommand{\0}{\ensuremath{\left|0\right\rangle}}
\newcommand{\1}{\ensuremath{\left|1\right\rangle}}
\newcommand{\2}{\ensuremath{\left|2\right\rangle}}
\newcommand{\proj}[1]{\ensuremath{\left|#1\right\rangle\left\langle #1\right|}}

\def\ket#1{\left|#1\right\rangle}

\newcommand{\A}{{\rm A}}
\newcommand{\B}{{\rm B}}
\newcommand{\AB}{{\rm AB}}

\makeatletter
\newcommand*{\addFileDependency}[1]{
  \typeout{(#1)}
  \@addtofilelist{#1}
  \IfFileExists{#1}{}{\typeout{No file #1.}}
}
\makeatother

\begin{document}

\title{Complete unitary qutrit control in ultracold atoms}

\author{Joseph Lindon}
\affiliation{Department of Physics, University of Alberta, Edmonton, Alberta, Canada}
\author{Arina Tashchilina}
\affiliation{Department of Physics, University of Alberta, Edmonton, Alberta, Canada}
\author{Logan W. Cooke}
\affiliation{Department of Physics, University of Alberta, Edmonton, Alberta, Canada}
\author{Lindsay J. LeBlanc}
\affiliation{Department of Physics, University of Alberta, Edmonton, Alberta, Canada}

\date{\today}

\begin{abstract}

Physical quantum systems are commonly composed of more than two levels and offer the capacity to encode information in higher-dimensional spaces beyond the qubit, starting with the three-level qutrit. 
Here, we encode neutral-atom qutrits in an ensemble of ultracold \Rb{} and demonstrate arbitrary single-qutrit SU(3) gates.  We generate a full set of  gates using only two resonant microwave tones, including synthesizing a gate that  effects a  coupling between the two disconnected levels in the three-level $\Lambda$-scheme. Using two different gate sets, we implement and characterize the Walsh-Hadamard Fourier transform, and find similar final-state fidelity and  purity from both approaches. This work establishes the ultracold neutral-atom qutrit as a promising platform for qutrit-based quantum information processing, extensions to $d$-dimensional qudits, and explorations in multilevel quantum state manipulations with nontrivial geometric phases.
\end{abstract}

\maketitle

\paragraph{Introduction.} The conventional paradigm for universal quantum computing makes use of two-level qubits, but higher-dimensional quantum systems offer considerable advantages. Logical operations and information storage using three-level systems -- ``qutrits'' -- in a larger Hilbert space give way to algorithms that can be more efficient~\cite{Wang2020} and allow for more complex entanglement than qubits~\cite{Caves2000}. Qutrits~\cite{Dogra2014} and higher $d$-dimensional qudits~\cite{Gedik2015} are valuable as a quantum resources for speed-up and for improved cryptographic security in transmission over quantum networks~\cite{Brus2002,Molina-Terriza2005,Ivanov2012}. Additionally, using a third temporary-state level during the implementation of a qubit gate can significantly improve fidelities and reduce circuit complexity~\cite{Lanyon2009,Ivanov2012,Bocharov2017,Gokhale2020}.

While qubits are readily simulated by the polarization of classical light, qutrits offer an additional complexity that reflects their inherently quantum properties~\cite{Dogra2014}.
Many physical platforms have served as host to qutrits~\cite{Klimov2003, McHugh2005,Wang2020}, including photonic systems~\cite{Bogdanov2004,Molina-Terriza2005,Groblacher2006,Lanyon2008,Schaeff2015,Babazadeh2017,Lu2020}, NMR ensembles~\cite{Dogra2014}, superconducting quantum circuits~\cite{Bianchetti2010,Danilin2018,Yurtalan2020, Kononenko2021, Blok2021,Morvan2021}, and trapped ions~\cite{Klimov2003,Randall2015,Baekkegaard2019,Low2020,Ringbauer2021}. 
In contrast to qubit operations, single \emph{qutrit} gates are represented by 3$\times$3 unitary matrices in the group SU(3), generated by the eight Gell-Mann matrices $\hat \lambda_i$~\cite{Caves2000}. Experimental qutrit demonstrations to date realized arbitrary operations by decomposing the desired transformation into three SU(2) operations followed by a diagonal phase gate~\cite{Klimov2003,Vitanov2012,Yurtalan2020,Kononenko2021}. In many physical systems,  two of the three SU(2) couplings are accessible, but selection rules dictated by parity mean that the third is less convenient or even forbidden.  To overcome this, two resonant couplings on the accessible transitions can be combined as a simultaneous dual-tone operator to synthesize the third coupling~\cite{Klimov2003}.

\begin{figure}[t!]
\includegraphics{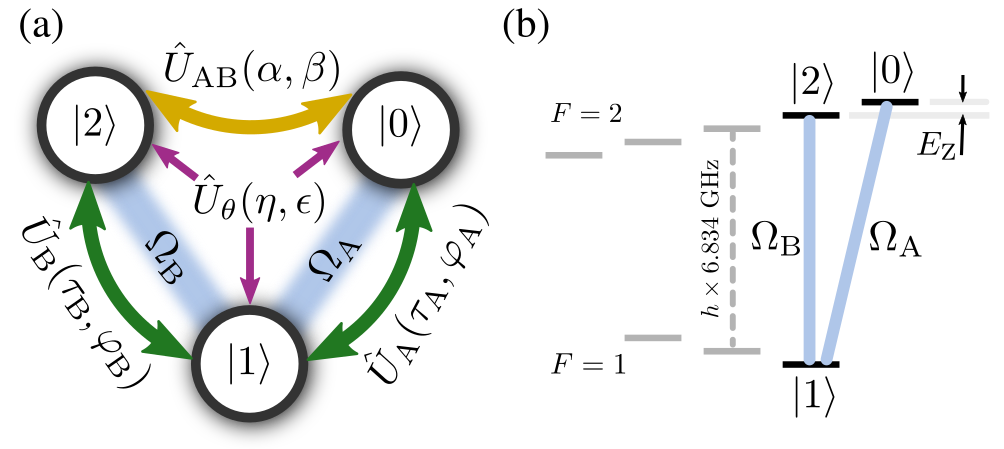}
\caption{(a) Schematic representation the three qutrit levels $\{\ket{0}, \ket{1}, \ket{2}\}$, where direct couplings ($\Omega_{\rm A}$ and $\Omega_{\rm B}$) exist only between $\ket{0}\leftrightarrow\ket{1}$ and $\ket{1}\leftrightarrow\ket{2}$, and are associated with unitary operations $\hat{U}_{\rm A}$
and $\hat{U}_{\rm B}$.  With a dual-tone operator, coupling between $\ket{0}\leftrightarrow\ket{2}$ is possible via $\hat{U}_{\rm AB}$; an additional phase-only operator $\hat{U}_\theta$ can be applied across all levels. (b) Energy level diagram of the $^{87}$Rb ground-state manifold, with levels $\ket{0} \rightarrow \ket{F=2, m_F = 2}$, $\ket{1} \rightarrow \ket{F=1, m_F = 1}$, and $\ket{2} \rightarrow \ket{F=2, m_F = 1}$ connected by two magnetic dipole transitions at microwave frequencies, whose energy difference is controlled by the Zeeman splitting, $E_{\rm Z}$.
}
\label{fig:schematic}
\end{figure}

In this work, we demonstrate neutral-atom qutrits  in ultracold ensembles, taking advantage of the well-defined, long-lived, and readily controlled hyperfine transitions in the alkali metals. 
We synthesize arbitary single-qutrit gates using two different schemes, both of which require only two couplings in the three-state manifold (Fig.~\ref{fig:schematic}).  In one of these approaches~\cite{Klimov2003}, we show an effective coupling between two levels that are not simply connected by microwave fields, establishing a comprehensive set of two-state operations within the three-state manifold.
{Our work specifically isolates and controls individual couplings within a qutrit with simple unitary decompositions that can be applied, in principle, to other qutrit platforms. This work builds beyond previous experiments~\cite{Chaudhury2007,Anderson2015} which use highly-parameterized control waveforms and gradient ascent to effect a target unitary on the space of one or more hyperfine manifolds.} 
Ensemble systems like ours provide a testbed for general quantum information processing with atoms~\cite{Barrett2010,Saffman2016}, though progress in atom-based quantum information processing with 
neutral-atom arrays~\cite{Saffman2016,Olmschenk2010,Henriet2020,Bluvstein2022} is rapidly advancing. 
Adapting the ensemble approach described here to address individual atoms could be achieved through site-specific gradients, focussed light shifts, or phase control~\cite{Wang2015,Wang2016b,Wang2020d,Steinert2022}, opening up the advantages of qutrits across atomic quantum computing platforms.  More broadly, the approach to quantum state control demonstrated here will provide tools for exploring fundamental operations on multilevel systems, including engineering nontrivial geometric phases in atomic systems~\cite{Novicenko2019,Chen2020,Sugawa2021}.

\noindent\paragraph{Single qutrit gates.}
A qutrit $|\psi\rangle=\sum_{k=0}^2 c_k e^{i\phi_k}\ket{k}$ is defined by the parameters $\{c_k\}$ and $\{\phi_k\}$, four of which are independent after considering normalization and global phase invariance. A single-qutrit gate, part of SU(3), is defined by eight independent parameters~\cite{Vitanov2012}. In a system with only two (complex) couplings [Fig.~\ref{fig:schematic}(a)], sequential operations can be applied to span all eight parameters, followed by a two-parameter relative phase adjustment, $\hat{U}_\theta(\eta,\epsilon)$.
In one approach~\cite{Dogra2014,Kononenko2021,Morvan2021}, a general single-qutrit gate is implemented with single-tone operators, such as
\begin{align}
\hat U^{\rm I}_{\rm gen.} = \hat U_\theta(\eta,\epsilon) \hat U_{\rm B}(\tau_{\rm B2},\varphi_{\rm B2}) \hat U_{\rm A}(\tau_{\rm A1},\varphi_{\rm A1})\hat U_{\rm B}(\tau_{\rm B1},\varphi_{\rm B1}),\nonumber
\end{align}
where $\tau_{\rm Xi} = |\Omega_{\rm X}| t/2$ and $\varphi_{\rm Xi} = \mathrm{arg}(\Omega_{\rm X})$ are the pulse areas and coupling phases for transitions ${\rm X}=\{\rm A,B\}$ with parameter index $\mathrm{i}$, and $\{\eta,\epsilon\}$ are relative phase adjustments~\cite{SupplMat}.

In a second approach~\cite{Klimov2003,SupplMat}, a dual-tone operator is used to simultaneously drive both transitions  A and B to synthesize a third coupling $\hat U_{\rm AB}(\alpha,\beta)$  between states $\ket{0}\leftrightarrow\ket{2}$, where the operator duration is exactly $t_{\rm AB} ={2\pi}/{\sqrt{|\Omega_{\rm A}|^2+|\Omega_{\rm B}|^2}}$ and the parameters $\alpha$ and $\beta$ are derived from the complex couplings $\Omega_{\rm A,B}$~\cite{SupplMat}.  A general gate is implemented through a combination of operators such as
\begin{align}
\hat U^{\rm II}_{\rm gen.} = \hat U_\theta(\eta,\epsilon) \hat U_{\rm B}(\tau_{\rm B2},\varphi_{\rm B2}) \hat U_{\rm A}(\tau_{\rm A1},\varphi_{\rm A1})\hat U_{\rm AB}(\alpha,\beta).\nonumber
\end{align}
In both decompositions I and II, the first SU(2) operator $\hat U_{\rm AB,B}$ is designed such that $\hat U_{\rm{gen.}}\hat U^\dagger_{\rm AB,B}$ has one off-diagonal matrix element of zero. The second pulse $\hat U_{\rm A}$ zeroes three more off-diagonal matrix elements in $\hat U_\mathrm{\rm gen.}\hat U^\dagger_{\rm AB,B} \hat U^\dagger_{\rm A}$, and after the third only a diagonal phase remains which is implemented by $\hat U_\theta$~\cite{SupplMat}.

As a particular example of a single-qutrit gate that uses control over all eight parameters, we consider the Walsh-Hadamard gate, which is the single-qutrit Fourier transform~\cite{Klimov2003}, 
\begin{align}
    \hat F = \frac{1}{\sqrt{3}}
    \begin{pmatrix}
        1&1&1\\
        1&e^{i \frac{2\pi}{3}} & e^{-i \frac{2\pi}{3}} \\
        1&e^{-i \frac{2\pi}{3}} & e^{i \frac{2\pi}{3}}
    \end{pmatrix}.
    \label{eq:F}
\end{align}
This transform has broad applications, such as Shor's algorithm~\cite{Vitanov2012}, error correction~\cite{Yurtalan2020}, implementing the SWAP gate, and the Bernstein-Vazirani algorithm~\cite{Wang2020}.  In the single- and dual-tone operator decompositions described above, the Fourier transforms are implemented as 
\begin{align}
\hat F^{\rm I} &=  e^{i\pi/6}\hat U_\theta(-\tfrac{\pi}{6},-\tfrac{\pi}{6}) 
            \hat U_{\rm B}\left(\tfrac{5\pi}{4},\tfrac{\pi}{2}\right)  
            \hat U_{\rm A}\left(\tau_{\rm A}^{\rm I}, \pi\right)
            \hat U_{\rm B}\left(\tfrac{\pi}{4},0\right),
            \nonumber\\
\hat F^{\rm II} &= i\hat U_\theta\left(\tfrac{\pi}{3},-\tfrac{\pi}{2}\right) 
            \hat U_{\rm A}\left(\tfrac{\pi}{4},\tfrac{\pi}{6}\right)
            \hat U_{\rm B}\left(\tau_{\rm B}^{\rm II}, \tfrac{\pi}{3}\right) 
            \hat U_{\rm AB}\left(\tfrac{\pi}{4},-\tfrac{2\pi}{3}\right), 
            \nonumber
\end{align}
where $\tau_{\rm A}^{\rm I} = \arccos(-1/3)/2$ and $\tau_{\rm B}^{\rm II} = \pi + \arctan(1/\sqrt{2})$.

We implement both the single- and dual-tone Fourier transform operators to  experimentally explore ultracold atomic ensembles as a platform for qutrit operations, and to compare the two approaches to qutrit operators in terms of final state fidelity and purity.

\paragraph{Experimental Methods.}
In our experiments, we prepare ultracold ensembles of \Rb{} atoms, manipulate the internal electronic states [Fig.~\ref{fig:schematic}(b)] using resonant microwave pulses, and measure the results by analyzing the final state via absorption imaging.  The ensemble approach allows us to perform simultaneous experiments on a large number ($\sim 10^5$) of identical atoms using  spatially uniform fields, and the measurement over the entire ensemble gives a statistical measure of the final state in a single experimental run. {The atoms remain coherent for well over \SI{1}{ms}, ensuring enough time to complete all operations in this protocol without dephasing or decoherence.}  

\begin{figure}[t!]
\includegraphics{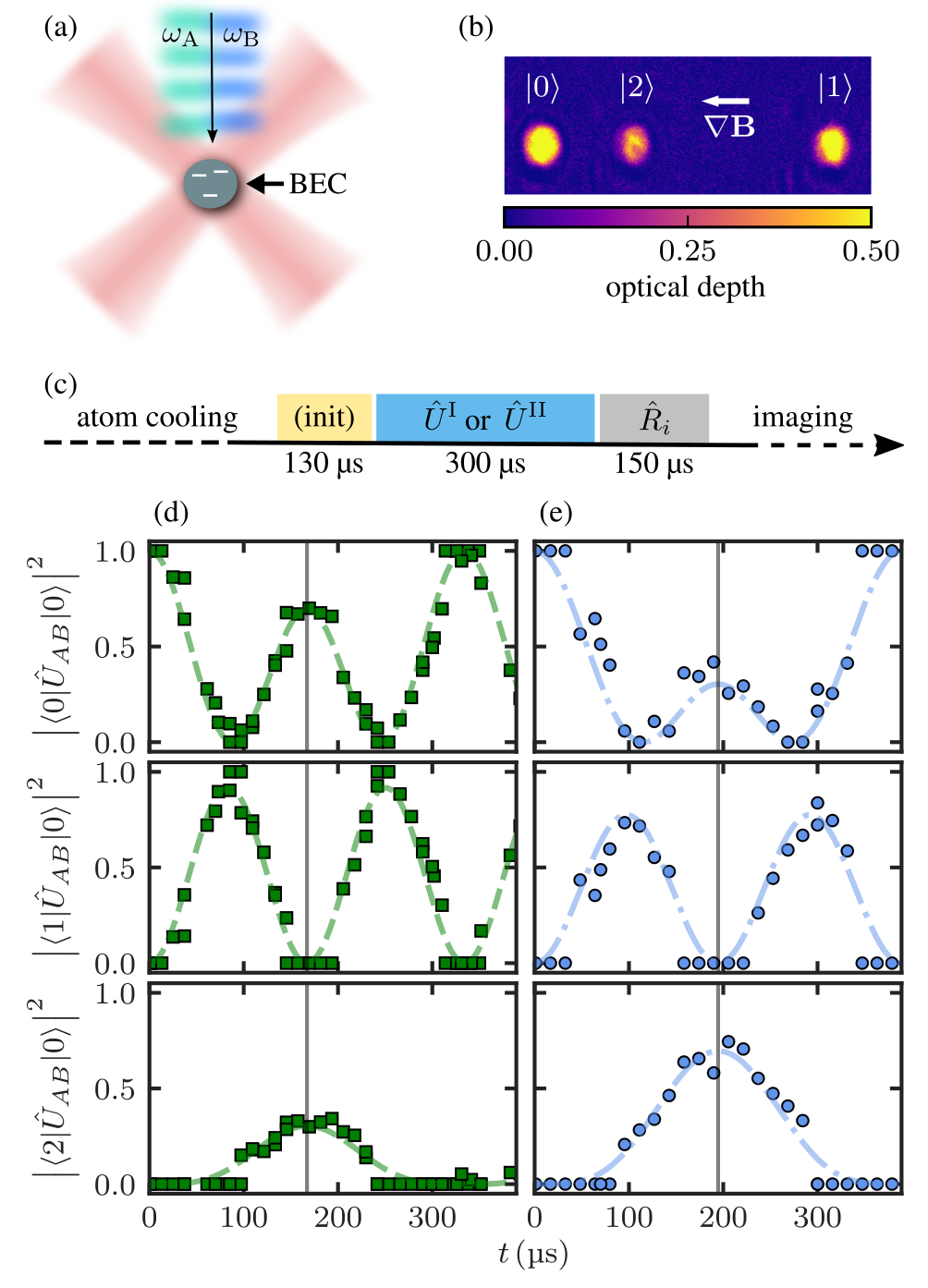}
\caption{
(a) A BEC of \Rb{} is trapped at the intersection of two optical dipole beams. Resonant microwave radiation with frequencies $\omega_{\rm A}$ and/or $\omega_{\rm B}$ drives transitions between levels.  
(b) A magnetic field gradient $\mathbf{\nabla B}$ is applied to the atoms during time-of-flight, before absorption imaging, spatially separating atomic levels according to their magnetic moments, $\mu_i$, where $\mu(\0) = \mu_{\rm B}$, $\mu(\1) = -\mu_{\rm B}/2$, and $\mu(\2) = \mu_{\rm B}/2$, and $\mu_{\rm B}$ is the Bohr magneton. The colour map represents an absorption image after \SI{25}{\milli\second} time of flight, with color bar indicating the optical depth of the atoms in clouds associated with each level. (c) Timing sequence for cooling, state initialization, qutrit gates $\hat U^{\rm I, II}$, and tomography rotations $\hat{R}_i$. (d,e) Example calibration data for dual-tone operators $\hat{U}_{\rm AB}(0.19\pi, 0)$ in (d) and $\hat{U}_{\rm AB}(0.31\pi, 0)$ in (e). Vertical gray lines indicate the operator time $t_{\rm AB}$ for which the intermediate-state \1 population/amplitude is zero. }
\label{fig:Rb}
\end{figure}

In particular, we prepare $10^5$ Bose-Einstein condensed \Rb{} atoms in an  {actively stabilized} optical dipole trap. 
The ground $5{}^{2}S_{1/2}$ state of \Rb{} has two hyperfine levels, $F=1$ and $F=2$,  separated by $\omega_0=2\pi\times \SI{
6.835
}{\giga\hertz}$,  each of which hosts $2F+1$ Zeeman sub-levels, which are energetically split by  $E_{\rm Z}\approx h \times \SI{1.7}{\mega\hertz}$ in a weak magnetic field. Using optical pumping, we initialize all atoms in the spin-polarized input state $|\psi_{\rm in}\rangle= \ket{0}\equiv \ket{F=2, m_F=2}$. {All microwave operations are triggered to begin at the same phase of a 60 Hz line oscillation.} Optionally, we  use $\pi$-pulse operations ($\hat U_{\rm A}$ or $\hat U_{\rm B}\hat U_{\rm A}$ pulses) to prepare initial states $\ket{\psi_{\rm in}} = \ket{1} $ or $\ket{2}$.  In this work, we use only microwave couplings between states of different $F$ to complete operations, avoiding the radiofrequency couplings between levels in the same $F$ manifold, which would allow direct \0 to \2 coupling.  By doing this, we avoid simultaneous couplings between all $m_F$ states in that manifold, due to the degeneracy of the Zeeman  transitions in a weak magnetic field.  The microwave transitions we use are, in contrast,  nondegenerate, even in weak magnetic fields.

Amplitude- and phase-controlled microwave signals, tuned near $\omega_0$, are resonant with the ``A''  $\ket{0}\leftrightarrow\ket{1}\equiv\ket{F = 1, m_F = 1}$
and ``B'' $\ket{1}\leftrightarrow\ket{2}\equiv\ket{F = 2, m_F = 1}$ transitions with typical Rabi frequencies $|\Omega_{\rm A,B}| \sim 2\pi \times 2$~kHz.  These two microwave tones, timed appropriately, effect the unitary operations $\hat U_{\rm A}$ or $\hat U_{\rm B}$ when used alone, and $\hat U_{\rm AB}$ when driven together~\cite{SupplMat}. Figures~\ref{fig:Rb}(c,d) show the dual-tone operator $\hat U_{\rm AB}$ acting on initial state $\ket{\psi_{\rm in}}= \ket{0}$ for varying pulse times: at the operator time $t_{\rm AB}$, the results show the populations distributed between $\ket{0}$ and $\ket{2}$ only.  This  distribution is controlled by the parameters $\alpha$ and $\beta$, which themselves depend on the amplitudes and phases of $\Omega_{\rm A}$ and $\Omega_{\rm B}$.

To effect a general single-qutrit gate, up to three of these operators are applied to the system, along with relative phase control~\cite{SupplMat}.
To decipher the amplitude and phase information of the resulting output state qutrits $\ket{\psi_{\rm out}}$, we perform tomography by applying rotation operations to the system (see below for details).  After the operations are complete, the atoms are released from the trap and the populations $|c_i|^2$ of each level in the qutrit are measured via absorption imaging: the three levels are spatially separated by a  Stern-Gerlach magnetic field gradient in time-of-flight, and counted simultaneously in a single absorption image~[Fig.~\ref{fig:Rb}(b)].

\paragraph{Tomography.}
To determine the full effect of the single-qutrit gates, we perform quantum state tomography on the final states $|\psi_{\rm out}\rangle$ for each of $\hat U_{\rm gen}^{\rm I}$ and $\hat U_{\rm gen}^{\rm II}$ acting on three orthogonal input states, $\ket{\psi_{\rm in}} = \{\ket{0},\ket{1},\ket{2}\}$.
Eight linearly independent projections of the qutrit state are required to fully characterize its density matrix $\hat \rho=\tfrac{\hat \openone}{3} + \tfrac{1}{3}\sum_{k=1}^8 \langle\hat\lambda_k\rangle\hat\lambda_k$~\cite{Caves2000}.

To perform the tomography, read-out operators $\hat R_i$ (Table~\ref{tab:tomography}) are applied before  projective measurement.  Table~\ref{tab:gm-tomo} shows how the orthogonal Gell-Mann matrices can be constructed from these projections. The usual projection operation $\langle \psi_{\rm out} | \chi\rangle\langle\chi|\psi_{\rm out}\rangle$ extends to $\Tr(|\chi\rangle\langle\chi|\hat \rho_{\rm out} )$ when a density matrix $\hat \rho_{\rm out}$ is substituted for the pure state $\ket{\psi_{\rm out}}$. We measure three projections for each of the six read-out operators, $\langle\hat R^\dagger_i\proj{0}\hat R_i\rangle$, $\langle\hat R^\dagger_i\proj{1}\hat R_i\rangle$, and $\langle\hat R^\dagger_i\proj{2}\hat R_i\rangle$, and in total, we measure $6\times 3$ projections, which are not all linearly independent.

\begin{table}[t!]

\caption{Read-out operators used for quantum state tomography. They are all $\pi/2$-area pulses.}
\begin{ruledtabular}
\begin{tabular}{c c c c}
    
    $\hat R_i$ & Basis & Phase $\phi$ &  Matrix form \\ \hline

    $\hat R_1$ & $\{\0,\1\}$ & $0$ &  $\frac{1}{\sqrt{2}} \begin{pmatrix}
        1&-1&0\\ 1 & 1 & 0 \\ 0 & 0 & \sqrt{2} \end{pmatrix}$\\
    
    $\hat R_2$ & $\{\0,\1\}$ & $\frac{\pi}{2}$ &  $\frac{1}{\sqrt{2}} \begin{pmatrix}
        1&-i&0\\ -i & 1 & 0 \\ 0 & 0 & \sqrt{2} \end{pmatrix}$\\
    
    $\hat R_3$ & $\{\1,\2\}$ & $0$ &  $\frac{1}{\sqrt{2}} \begin{pmatrix} \sqrt{2} & 0 & 0 \\ 0 & 1 & 1 \\ 0 & -1 & 1 \end{pmatrix}$ \\
    
    $\hat R_4$ & $\{\1,\2\}$ & $\frac{\pi}{2}$ & $\frac{1}{\sqrt{2}} \begin{pmatrix} \sqrt{2} & 0 & 0 \\ 0 & 1 & -i \\ 0 & -i & 1 \end{pmatrix}$\\

    $\hat R_5$ & $\{\0,\2\}$ & $0$ & $\frac{1}{\sqrt{2}} \begin{pmatrix}
        1 & 0 & -1 \\ 0 & -\sqrt{2} & 0 \\ -1 & 0 & -1 \end{pmatrix}$\\
    
    $\hat R_6$ & $\{\0,\2\}$ & $\frac{\pi}{2}$ & $\frac{1}{\sqrt{2}}\begin{pmatrix}
        1 & 0 & -i \\ 0 & -\sqrt{2} & 0\\ i & 0 & -1
    \end{pmatrix}$
\end{tabular}
\end{ruledtabular}

\label{tab:tomography}
\end{table}

\begin{table}[tb!]
\caption{Demonstration of the Gell-Mann matrices constructed using read-out operators $\hat R_i$.}
\begin{ruledtabular}
\begin{tabular}{c c c}
$\hat \lambda_i$ & Construction \\ \hline
$\hat \lambda_1$ & $\hat R_1^\dagger\left(\proj{1}-\proj{0}\right)\hat R_1$ \\
$\hat \lambda_2$ & $\hat R_2^\dagger\left(\proj{0}-\proj{1}\right)\hat R_2$ \\
$\hat \lambda_3$ & $\hat R^\dagger_3 \proj{2}\hat R_3 - \hat R^\dagger_5 \proj{1} \hat R_5$\footnote{One of 4 possibilities shown}\\
$\hat \lambda_4$ & $\hat R^\dagger_5\left(\proj{2}-\proj{0}\right)\hat R_5$ \\
$\hat \lambda_5$ & $\hat R^\dagger_6\left(\proj{0}-\proj{2}\right)\hat R_6$ \\
$\hat \lambda_6$ & $\hat R^\dagger_3\left(\proj{1}-\proj{2}\right)\hat R_3$ \\
$\hat \lambda_7$ & $\hat R^\dagger_4\left(\proj{1}-\proj{2}\right)\hat R_4$ \\
$\hat \lambda_8$ & $\left(\hat R^\dagger_3\proj{0}\hat R_3 + \hat R^\dagger_5 \proj{1}\hat R_5 - 2\hat R^\dagger_1 \proj{2} \hat R_1\right)/\sqrt{3}$\footnote{One of 8 possibilities shown}
\end{tabular}
\end{ruledtabular}
\label{tab:gm-tomo}
\end{table}

An iterative maximum likelihood technique~\cite{Lvovsky2004} allows us to estimate the density matrix $\hat \rho$ while maintaining the condition $\Tr(\hat \rho) = 1$. Using the function 
\begin{align}
    \hat Q(\hat \rho) &= \sum_{i}\sum_{j=0,1,2} \frac{f^i_j}{\Tr\left(\hat R_i^\dagger|j\rangle\langle j|\hat R_i\hat \rho\right)}\hat R_i^\dagger|j\rangle\langle j|\hat R_i ,
    \label{eq:Q}
\end{align}
where $f^i_j$ is the cloud fraction found experimentally in eigenstate $|j\rangle$ after the read-out operator $\hat R_i$ is applied, we iterate through
\begin{align}
    \hat \rho_{i+1} = \frac{\hat Q(\hat \rho_i) \hat \rho_i\hat Q(\hat \rho_i)}{\Tr\left(\hat Q(\hat \rho_i) \hat \rho_i\hat Q(\hat \rho_i)\right)} 
\end{align}
until convergence, having begun with the maximally mixed density matrix $\hat \rho_0=\hat \openone /3$ (Ref.~\cite{Caves2000}).

\paragraph{Results.}

\begin{figure}[t!]
    \centering
    \includegraphics{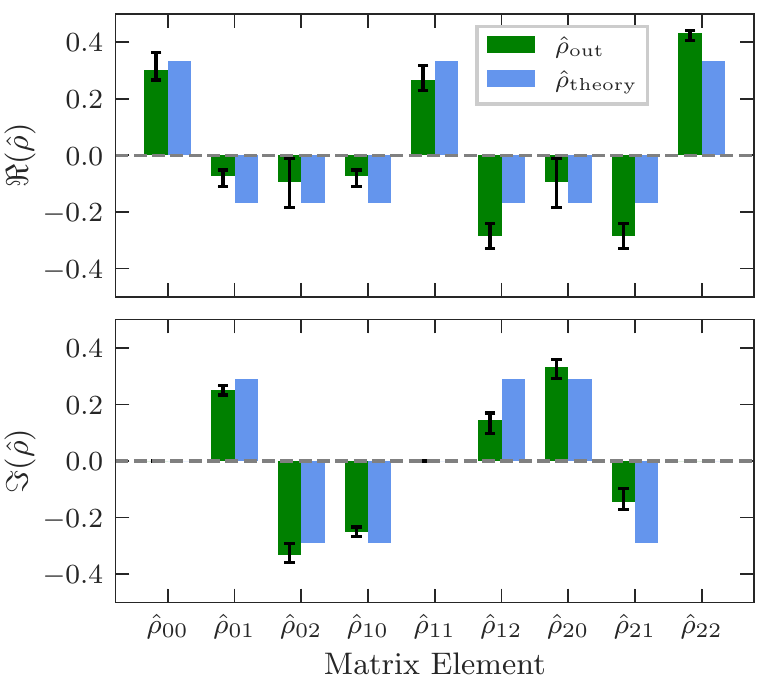}
    \caption{Reconstructed density matrix elements from the measured values $\hat\rho_{\rm out}$ (green, left bars) and for the modelled, expected density matrix  $\hat\rho_{\rm theory}$ (blue, right bars) for the Fourier operation $\hat F^{\rm II} \0$.  The real (upper) $\Re(\hat\rho)$ and imaginary (lower) $\Im(\hat\rho)$  parts of these elements are shown. Error bars show the range of each matrix element across all $N=10$ tomographic measurements. As a Hermitian operator, the main diagonal of $\hat \rho$ is real.}
    \label{fig:Fk0}
\end{figure}

We applied the  Walsh-Hadamard decompositions $\hat F^{\rm I}$ and $\hat F^{\rm II}$ on three input computational basis states \0, \1, and \2 and characterize the output state's density matrix,  $\hat \rho_{\rm out}$.
Figure~\ref{fig:Fk0} shows one such density-matrix reconstruction, after applying $\hat F^{\rm II}$ to the state \0.
For each  tomographically measured $\hat \rho_{\rm out}$ (resulting from one of $\hat{F}^{\rm I,II}$ acting on an input basis state $\ket{n}$), we calculate the quantum state fidelity~\cite{Jozsa94}
\begin{equation}
    \mathcal{F}(\hat F, \ket{n})=\langle n  | \hat F^\dagger \hat \rho_\mathrm{out} \hat F|  n\rangle
    \label{eq:fidelity}
\end{equation}
and  purity $ \mathcal{P}(\hat \rho_{\rm out}) = \Tr(\hat\rho^2_\mathrm{out})$. 
Table~\ref{tab:rho-fidelities} shows that the single qutrit gates operate as expected for all input basis states, and that the two decompositions produce states with similar fidelities and purities.

\begin{table}[tb!]
    \centering
    \caption{Fidelity and purity found by maximum likelihood estimation after state tomography for  two decompositions of the single-qutrit Fourier transform. Errors shown are the standard deviation of values across the $N$ tomographic measurements. State preparation and measurement errors are not removed.}
    \begin{ruledtabular}
    \begin{tabular}{c c c c c c}
        Operator & $\ket{\psi_{\rm in}}$& $N$ & $\mathcal{P}$ &  $\mathcal{F}$ & $\mathcal{F}_\mathrm{pure}$\\ 
         
\hline\multirow{3}{*}{$\hat F^{II}$}
 & \0 & 10 & 0.93(4) & 0.91(2) & 0.95(2) \\
 & \1 & 10 & 0.92(3) & 0.91(2) & 0.96(2) \\
 & \2 & 10 & 0.90(6) & 0.86(4) & 0.92(2) \\
\hline\multirow{3}{*}{$\hat F^{I}$}
 & \0 & 15 & 0.90(8) & 0.92(6) & 0.98(3) \\
 & \1 & 12 & 0.88(8) & 0.87(5) & 0.95(3) \\
 & \2 & 16 & 0.89(4) & 0.89(7) & 0.95(7) \\
    \end{tabular}
    \end{ruledtabular}
    
    \label{tab:rho-fidelities}
\end{table}

\begin{figure}[tb!]
    \centering
    \includegraphics{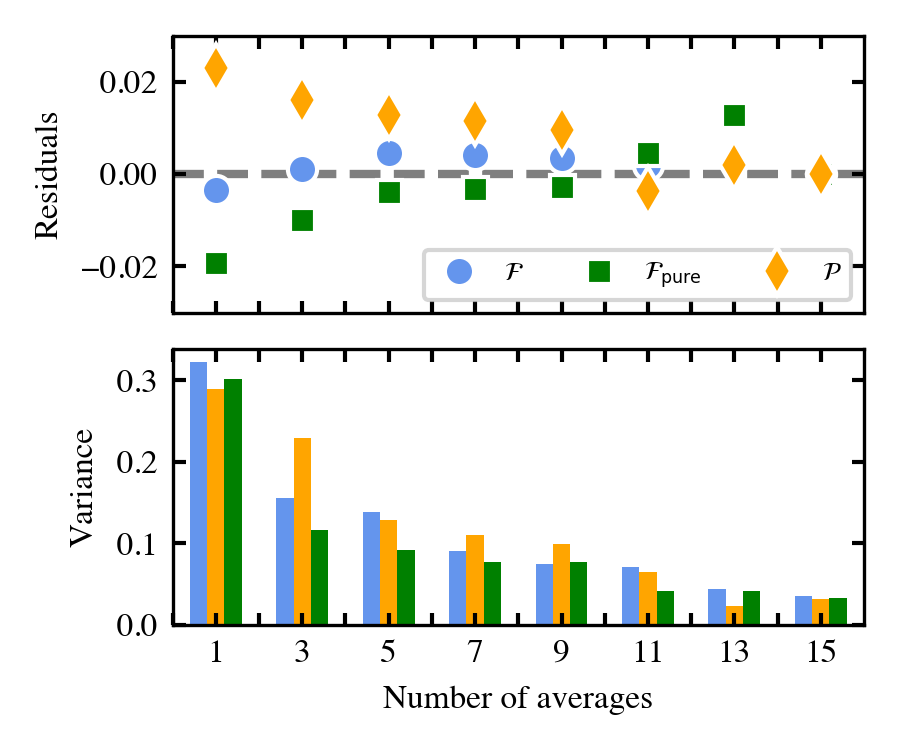}
    \caption{Gate quality metrics $\mathcal{F}$, $\mathcal{P}$, and $\mathcal{F}_{\rm pure}$ after averaging over $N$ experimental scans. The upper plot shows the residuals for results with respect to the average over $N = 15$ scans, and the lower plot shows the variance of the results (given as the variance between the maximum and the minimum of each metric): left blue bars: $\mathcal{F}$, middle yellow bars: $\mathcal{P}$, right green bars: $\mathcal{F}_{\rm pure}$. 
    While the variance reduces with more averaging, as expected, there is minimal change in the average value (seen via the residuals), suggesting that random errors from one experiment to the next is not a dominant error mechanism.}
    \label{fig:fidelity-avg}
\end{figure}

The ensemble approach taken in this work gives excellent statistics and provides a fast path towards calibrating the pulse areas and phases.  However, this approach is not without its limitations, which we see in the values for both the fidelity $\mathcal{F}$ and purity $\mathcal{P}$. After averaging each measurement over $N\ge 10$ experimental trials [Fig.~\ref{fig:fidelity-avg}], we find that variations in the results decrease with increasing $N$, but the average values remain similar to the single $N=1$ trial.  This indicates that shot-to-shot noise is not a significant contributor to the infidelity and impurity of these measures.  In contrast, we find that the impurity of the final states impacts the fidelity: if we assume the nearest pure state is achieved before calculating the fidelity using Eq.~\ref{eq:fidelity} by estimating that state as
\begin{equation}
    \hat \rho \rightarrow \frac{\hat \rho-\hat \openone/3}{\mathcal{P}(\hat \rho)} + \frac{\hat \openone}{3}.
    \label{eq:purity-recover}
\end{equation}  
we find  better $\mathcal{F}_\mathrm{pure}$ than $\mathcal{F}$, indicating that a significant part of the infidelity arises from loss of purity. 
Spatial inhomogeneities in the coupling fields or the environment experienced by atoms may cause position-dependent evolution and appear as a loss of ensemble purity.
Our analysis~\cite{SupplMat} suggests the dominant effect is dephasing of the qutrit due to  Stark-shift inhomogeneity in the optical dipole trap: atoms in the centre of the trap have a detuning on the order of \SI{400}{\hertz} relative to those at the edges, and because the atoms are cold and move little during the operation time, this difference does not average out over the course of a gate sequence. 
A probable source of the the remaining infidelity is imprecision in the calibration of  individual operators.

\paragraph{Discussion.}
We successfully demonstrated arbitrary SU(3) control in neutral alkali atoms using the Walsh-Hadamard (Fourier transform) single-qutrit gate, while implementing the resonant dual-tone operator  $\hat U_{\rm AB}$. We find that two different decompositions of arbitrary SU(3) gates using three SU(2) rotations result in comparable fidelities.
The dual-tone operator $\hat U_{\rm AB}$ is particularly useful for qutrit operations in platforms where one coupling is forbidden or inconvenient to use, not only in the ultracold atomic states used here, but also in systems such as superconducting qutrits~\cite{Yurtalan2020} and ions~\cite{Klimov2003}. {In our experiments, for example, we harnessed $U_\mathrm{AB}$ to perform each tomographic read-out operation in a single step, while previous works~\cite{Bianchetti2010,Yurtalan2020} have needed several pulses to prepare some projections.}

When decomposing SU($d$) operations into SU(2) steps, the number of operations scales quadratically with $d$. In the future, alternative approaches to  SU(3) operations could be implemented by decomposing via Householder reflections rather than SU(2) operations, which scales linearly with $d$ by applying dual-tone time-dependent pulses with detuning~\cite{Vitanov2012}.  Additionally, optimization using quantum control techniques is promising for increasing the speed and fidelity of qutrit and qudit operations~\cite{Randall2018,Omanakuttan2021}. Refs. \cite{Chaudhury2007,Anderson2015} have used phase-modulated signals for quantum control in ${}^{133}\mathrm{Cs}$ using $d=7$ and $d=16$. Using two fully independent transitions $A,B$ with their own modulations may lead to faster and more efficient optimized gates.

Looking forward, the two-qutrit operations necessary for universal quantum information processing~\cite{Luo2014,Low2020,Wang2020} will expand the scope of these initial demonstrations, whether in ensembles~\cite{Ebert2015} or single-atom arrays~\cite{Henriet2020}. Opportunities in broader areas, such as for holonomic computing~\cite{Zanardi1999,Feng2013,Zhou2017a,Danilin2018,Xu2020} will also take benefit from a comprehensive control over the multilevel state systems developed for qutrit and qudits.  While there is increasing potential for neutral atoms to serve in these roles, the general techniques for quantum state control over qudits apply across platforms, and developments across fields will rapidly accelerate this capabilities for all systems.

\begin{acknowledgments}
{We thank Barry C. Sanders for useful discussions.  This work was supported by the University of Alberta; the Natural Sciences and Engineering Research Council, Canada (Grants No. RGPIN-2021-02884 and No. CREATE-495446-17);  the Alberta Quantum Major Innovation Fund; Alberta Innovates; the Canada Foundation for Innovation, and the Canada Research Chairs  (CRC) Program.
We gratefully acknowledge that this work was performed on Treaty 6 territory, and as researchers at the University of Alberta, we respect the histories, languages, and cultures of First Nations, M\'etis, Inuit, and all First Peoples of Canada, whose presence continues to enrich our vibrant community.}

\end{acknowledgments}


\providecommand{\noopsort}[1]{}\providecommand{\singleletter}[1]{#1}%
%



\newpage
\onecolumngrid

\setcounter{equation}{0}
\setcounter{figure}{0}
\setcounter{table}{0}
\setcounter{page}{1}
\setcounter{section}{0}
\renewcommand{\theequation}{S\arabic{equation}}
\renewcommand{\thefigure}{S\arabic{figure}}
\renewcommand{\thetable}{S\arabic{table}}


{\large \bf Supplementary material: Complete unitary qutrit control in ultracold atoms}\\
\vspace{-10pt}
\begin{center}
 Joseph Lindon, Arina Tashchilina, Logan W. Cooke, and Lindsay J. LeBlanc \\
\emph{Department of Physics, University of Alberta, Edmonton, Alberta, T6G 2E1, Canada}
\end{center}

\section{Operator Derivations}
Here we provide details of the the operators necessary for the SU(3) decompositions $\hat U^{\rm I}$ and $\hat U^{\rm II}$. 

\subsection{Derivation of Coupling Operators}
The coupling operators $\hat U_{\rm A}$, $\hat U_{\rm B}$, and $\hat U_{\rm AB}$ are produced by driving resonant microwave couplings between states $\{\0, \1\}$ and $\{\1, \2\}$, as shown in Fig.~\ref{fig:Rb}(a).
In general, when such a resonant dual-tone field is applied to a three-level system, the lab-frame Hamiltonian in the basis $\{\0, \1, \2\}^T$ is
\begin{widetext}
\begin{align}
    \hat H = \hbar \begin{pmatrix}
        \omega_\A & |\Omega_\A| \sin\left(\omega_\A t+\phi_\A\right) & 0\\
        |\Omega_\A| \sin\left(\omega_\A t+\phi_\A\right) & 0 & |\Omega_\B| \sin\left(\omega_\B t+\phi_\B\right)\\
        0 & |\Omega_\B| \sin\left(\omega_\B t+\phi_\B\right) & \omega_\B
    \end{pmatrix},
\end{align}
\end{widetext}
where $\omega_{\rm X}$, $\phi_{\rm X}$ and $\Omega_{\rm X}$ are the resonant frequency, phase, and Rabi frequency for the transition ${\rm X} \in \{\A, \B\}$ shown in Fig.~\ref{fig:schematic}.
A useful rotating frame for this analysis is generated by the rotation
\begin{align}
    \hat W = \begin{pmatrix}
        e^{i\omega_\A t} & 0 & 0 \\
        0 & 1 & 0\\
        0 & 0 & e^{i\omega_\B t}
    \end{pmatrix}.
\end{align}
After transforming the Hamiltonian into this frame $\hat{\tilde H} = \hat W^\dagger \hat H \hat W + i\hbar (\partial_t \hat W)\hat W^\dagger$ and applying the rotating wave approximation, the new effective Hamiltonian is
\begin{align}
    \hat{\widetilde H} = \frac{i\hbar}{2}\begin{pmatrix}
        0 & -\Omega_\A & 0\\
        \Omega_\A^* & 0 & \Omega_\B^* \\
        0 & -\Omega_\B & 0
    \end{pmatrix}.
    \label{eq:DualRWA}
\end{align}
where $\Omega_X = |\Omega_X| e^{i\phi_X}$ is the complex coupling parameter. A single-qutrit gate operation performed by applying this Hamiltonian for some duration $t$ is
\begin{align}
    \hat U &= \exp(-i\hat {\widetilde H} t /\hbar).\label{eq:U}
\end{align}

As one ingredient in our decomposition, we consider  SU(2) coupling gates realized by applying a single tone field. If, for example, we impose the condition $\Omega_\B=0$ to eq. \eqref{eq:DualRWA} before using this Hamiltonian to calculate the evolution \eqref{eq:U} for some duration $t_A=2 \tau_\A / |\Omega_\A|$, we find a gate operating only on the $\0\leftrightarrow\1$ subspace
\begin{align}
    \hat U_A (\tau_\A, \phi_\A)= \begin{pmatrix}
        \cos\tau_\A &
        - e^{i \phi_\A} \sin\tau_\A & 0\\
        e^{-i\phi_\A} \sin\tau_\A & \cos\tau_\A & 0\\
        0 & 0 & 1
    \end{pmatrix},
\end{align}
which provides amplitude and phase control for the targeted coupling between \0$\leftrightarrow$\1. Similarly, for the condition $\Omega_\A=0$ and duration $t_\B=2\tau_\B / |\Omega_\B|$, we have
\begin{align}
    \hat U_{\rm B} (\tau_\B, \phi_\B) = \begin{pmatrix}
        1 & 0 & 0\\
        0 & \cos\tau_\B
        & e^{-i\phi_\B}\sin\tau_\B\\
        0 & -e^{i\phi_\B}\sin\tau_\B & \cos\tau_\B
    \end{pmatrix}
\end{align}
which achieves the same for for the targeted coupling between \1$\leftrightarrow$\2.

To implement $\hat U_{\AB}$, both coupling terms in \eqref{eq:DualRWA} are generally non-zero. If we restrict the pulse duration to $t_{\rm AB}=2\pi/\sqrt{|\Omega_{\rm \A}|^2+|\Omega_{\rm B}|^2}$ as was first explained in Ref.~\cite{Klimov2003}, we find the resonant dual-tone operator first shown in that paper, 
\begin{align}
    \hat U_{\rm AB} (\alpha, \beta) &=\begin{pmatrix}
        \cos(\alpha) & 0 & -e^{i\beta}\sin(\alpha)\\
        0 & -1 & 0\\
        -e^{-i\beta}\sin(\alpha) & 0 & -\cos(\alpha)
    \end{pmatrix},
    \label{eq:KlimovU1}
\end{align}
where to simplify the expression we have defined 
$\alpha=2 \arctan|\Omega_{\rm A}/\Omega_{\rm B}|$
and $\beta = \arg(\Omega_\mathrm{A}/\Omega_\mathrm{B})$.  Under this condition for the operator time, the coupling is directly between the states \0$\leftrightarrow$\2

In addition to the SU(2) couplings demonstrated here, a diagonal phase gate is required to span SU(3).  

\subsection{Virtual Phase Gates}
\label{sec:virtual-phase}
While Ref.~\cite{Klimov2003} suggests applying far-off resonance fields to perform the phase gate
\begin{equation}
    \hat U_\theta(\eta, \varepsilon) = \exp\begin{pmatrix}
        i\eta&0&0\\
        0&i\varepsilon & 0 \\
        0 & 0 & -i(\eta+\varepsilon)
    \end{pmatrix},
\end{equation}
recent qutrit experiments \cite{Kononenko2021, Yurtalan2020,Blok2021,Morvan2021} have implemented diagonal phase gates virtually, by manipulating proceeding control fields rather than by manipulating the atoms directly. As they are adjustments to control fields, these virtual phase gates are efficient and have zero duration~\cite{McKay2017}.

Any unitary operators $\hat U_a$ that are to be implemented after phase gate $\hat U_\theta$ are simply phase shifted by the transformation
\begin{equation}
    \hat{\tilde U}_a = \hat U_\theta^\dagger \hat U_a \hat U_\theta.
\end{equation}
This effectively delays application of the $\hat U_\theta$ operator until the end of the pulse sequence -- notice the location of $\hat U_\theta$ in the following operator sequence.
\begin{align}
\begin{split}
    \hat U_\mathrm{gen.} &= \hat U_{a_3} \hat U_{a_2} \hat U_{a_1} \hat U_\theta \hat U_b \\
    &= \hat U_\theta \hat U_\theta^\dagger \hat U_{a_3} \hat U_\theta \hat U_\theta^\dagger \hat U_{a_2} \hat U_\theta \hat U_\theta^\dagger \hat U_{a_1} \hat U_\theta \hat U_b\\
    &= \hat U_\theta \hat{\tilde U}_{a_3} \hat{\tilde U}_{a_2} \hat{\tilde U}_{a_1} \hat U_b.
\end{split}
\end{align}
Several virtual phase gates are easily combined by tracking the accumulated $\eta$ and $\varepsilon$ through the operator sequence.

The remaining final $\hat U_\theta$ operator never needs to be applied because after the last coupling operator (including tomography pulses), the state is projected via $\left|\langle n | \hat U_\mathrm{gen.} | \psi\rangle\right|^2$, and the phase information is destroyed. 

In our experiment, the read-out operators $\hat R_i$ are applied after after the diagonal phase gate $\hat U_\theta$, so these pulses are phase shifted in practice, and modified pulses $\hat{\tilde R}_i$ are generated: 
\begin{align}
\hat{\tilde R}_{1,2}(\phi) &= \hat R_{1,2}(\phi+\varepsilon-\eta),\\
\hat{\tilde R}_{3,4}(\phi) &= \hat R_{3,4}(\phi - \eta - 2 \varepsilon), \\
\hat{\tilde R}_{5,6}(\phi) &= \hat R_{5,6}(\phi-\varepsilon-2\eta).
\end{align}

\section{\label{sec:F}Fourier Transform}
The Fourier transform $\hat F$ does not have determinant 1, i.e. it is not a member of the group SU(3), however, $i\hat F$ is in the group, so the operation can be achieved up to a global phase. The decomposition $\hat F^{\rm I}$ differs from \eqref{eq:F} by the change of basis $\1\leftrightarrow\2$, and the decomposition $\hat F^{\rm II}$ differs from \eqref{eq:F} by the change of basis $\0\leftrightarrow\1$.

\section{Decomposition of Operators}
The supplementary material of Ref.~\cite{Kononenko2021} shows an algorithm for decomposition of any arbitrary qutrit gate $\hat U_\mathrm{gen.}$ into SU(2) steps using two single-tone couplings. We show  a  version of the procedure that works for either single or dual-tone decompositions here for clarity and because we suspect that exposition has typographical errors. A SU(2) coupling on the basis $\{\ket{m}, \ket{n}\}$ of area $\tau$ and phase $\phi$ can be expressed as

\begin{equation}
    R^{mn}(\tau, \phi) = \exp\left\{-i \frac{\tau}{2}\left[\cos(\phi)\sigma_x^{mn} + \sin(\phi)\sigma_y^{mn}\right]\right\}
\end{equation}
where $\sigma_x^{mn}=|m\rangle\langle n| + |n\rangle\langle m|$ and $\sigma_y^{mn}=i|n\rangle\langle m| -i|m\rangle\langle n|$. Each step of the rotation has
\begin{equation}
    \tau = 2\arcsin\sqrt{\frac{|\langle a|\hat U_k|m\rangle|^2}{|\langle a|\hat U_k|m\rangle|^2+|\langle a|\hat U_k|n\rangle|^2}}
\end{equation}
and 
\begin{equation}
    \phi = \frac{\pi}{2} + \arg\left(\langle a|\hat U_k|m\rangle\right) - \arg\left(\langle a | \hat U_k | n \rangle\right)
\end{equation}
where $\langle a|\hat U_k |m\rangle$ is a matrix element of $\hat U_k$ to be zeroed and $\hat U_k$ is the remaining remaining portion of $\hat U_\mathrm{gen.}$ to be implemented. The values of $|a\rangle$, $|m\rangle$, $|n\rangle$, and $\hat U_k$ to be used for each step of the decomposition are given in Table~\ref{tab:decomp}, and a Python implementation of the general decomposition is provided in listing~\ref{lst:decomp}.

\renewcommand{\arraystretch}{1.4}
\begin{table*}[tb!]
\caption{Parameters for Operator Decomposition}
\label{tab:decomp}
\begin{tabular*}{5in}{@{\extracolsep{\fill}}l c l l l}
\hline
\hline
Decomposition & Step, $k$ & $\hat U_k$ & $a$ & Coupling basis $\{\ket{m}, \ket{n}\}$ \\ \hline
\multirow{3}{*}{$\hat U^{\rm I}=\hat U_\theta \hat U_\A \hat U_\B \hat U_\A$} & 1 & $\hat U_\mathrm{gen.}$ & \2 & A: $\{\0, \1\}$ \\
& 2 & $\hat U_\mathrm{gen.}\hat U_1^\dagger$ & \2 & B: $\{\1, \2\}$ \\
& 3 & $\hat U_\mathrm{gen.}\hat U_1^\dagger\hat U_2^\dagger$ & \1 & A: $\{\0, \1\}$ \\ \hline
\multirow{3}{*}{$\hat U^{\rm II}=\hat U_\theta \hat U_\A \hat U_\B \hat U_{\AB}$} & 1 & $\hat U_\mathrm{gen.}$ & \2 & AB: $\{\0, \2\}$ \\
& 2 & $\hat U_\mathrm{gen.}\hat U_1^\dagger$ & \2 & B: $\{\1, \2\}$ \\
& 3 & $\hat U_\mathrm{gen.}\hat U_1^\dagger\hat U_2^\dagger$ & \1 & A: $\{\0, \1\}$\\
\hline
\hline
\end{tabular*}
\end{table*}

\section{Signal Generation}
To generate microwave signals with arbitrary phase, frequency, and amplitude control, we mix the output of an arbitrary waveform generator with a microwave signal. The microwave signal is detuned \SI{100}{\mega\hertz} below the hyperfine splitting of \Rb{}, and is produced by a BNC Model 845 Microwave Signal Generator. A Tektronix AWG5204 controlled by Python software produces signals at \SI[per-mode=symbol]{5}{\giga\siemens\per\second} which are pre-amplified before being combined with the microwave-frequency signal in a double-balanced mixer. The mixed signal is amplified by a \SI{25}{\watt} microwave amplifier before being directed towards the atoms through a waveguide. Higher Rabi frequencies would certainly be achievable with higher microwave power or a more focused beam.

\section{Error Mechanisms}
In this section we discuss and evaluate some mechanisms in our system that could cause the fidelity and purity errors from table \ref{tab:rho-fidelities} in the main text. {\color{blue} We note that dephasing is not a leading source of error, as our $T_2^*$ times are well over 1~ms.}

\subsection{Detuning}
When performing manipulations using our apparatus, we frequently find re-adjusting magnetic field biases necessary to keep the two generated frequencies resonant with the atomic frequencies $\omega_A$ and $\omega_B$. This field instability of order 30 minutes is a limitation of our system, and not a limitation of neutral atom computation in general. Because of concerns about the continued resonance of operations, in addition to frequent checks of the biases we average final populations for tomographic tests as discussed in the main text.

\begin{figure}[tb!]
    \centering
    \includegraphics{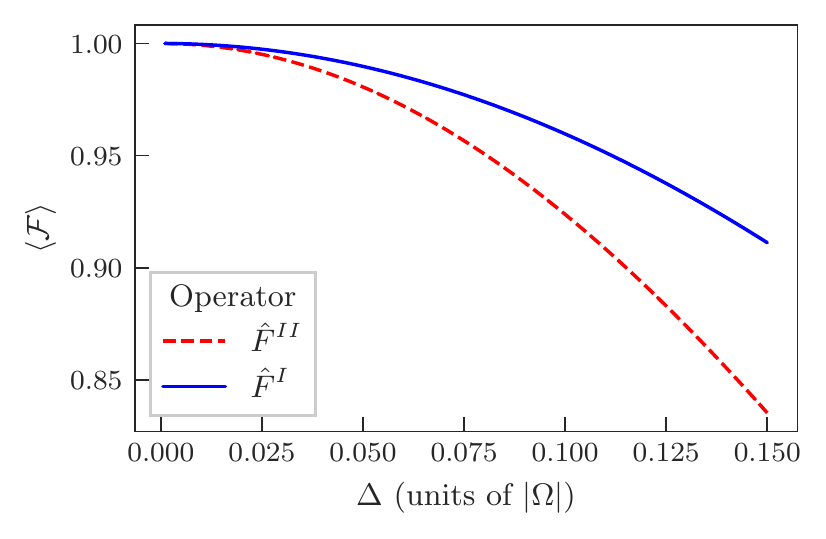}
    \caption{Simulated effect of detuning on the final fidelity measurement, averaged across the states \0, \1, and \2, where $\Omega^2=|\Omega_\A|^2+|\Omega_\B|^2$.}
    \label{fig:detuning_sim}
\end{figure}

We numerically simulate the operations $\hat F^{\rm I}$ and $\hat F^{\rm II}$, adding detuning deliberately,
\begin{widetext}
    
\begin{align}
    \hat H_\mathrm{detun.} = \hbar \begin{pmatrix}
        \omega_\A & |\Omega_\A|\sin((\omega_\A+\Delta_\A) t + \phi_\A) & 0\\
        |\Omega_\A|\sin((\omega_A+\Delta_\A) t + \phi_\A) & 0 & |\Omega_\B|\sin((\omega_\B+\Delta_\B) t + \phi_\B) \\
        0 & |\Omega_\B|\sin((\omega_\B+\Delta_\B) t + \phi_\B) & \omega_\B
    \end{pmatrix}
\end{align}
\begin{align}
    \hat{\widetilde H}_\mathrm{detun.} = \frac{i\hbar}{2} \begin{pmatrix}
        0 & -\Omega_\A e^{i\Delta_\A t} & 0\\
        (\Omega_\A e^{i\Delta_\A t})^* & 0 & (\Omega_\B e^{i\Delta_\B t})^* \\
        0 & -\Omega_\B e^{i\Delta_\B t} & 0
    \end{pmatrix},
\end{align}
\end{widetext}
and, setting $\Delta = \Delta_\A = -\Delta_\B$ for each of the decomposed SU(2) steps, generate the curves of figure \ref{fig:detuning_sim}.
If detuning applied equally to all atoms in the ensemble was a strong error mechanism in our system, we would expect to see lower $\mathcal{F}$ for the dual-tone $\hat F^{\rm II}$ than the single-tone $\hat F^{\rm I}$, and we would not expect to see a drop in $\mathcal{P}$. In our experiment, the fidelity for both decompositions is of a similar magnitude, and we observe a significant drop in $\mathcal{P}$, so this detuning model does not explain the main error mechanism present.

\subsection{Stark Shifts Caused by the Optical Dipole Trap}
During the application of microwave pulses for state manipulation, the atomic ensemble is held in a two-beam crossed optical dipole trap. This laser with approximately \SI{0.4}{\watt} of power holds atoms in place by creating a 3D Gaussian trap with depth \SI{6}{\micro\kelvin}. The potential experienced by the Bose-Einstein condensed atoms near its centre is approximated by a spherically symmetric harmonic oscillator with trap frequency $\omega_\mathrm{ho}\approx 2\pi\times \SI{100}{\hertz}$. The $10^5$ atoms are held within a Thomas-Fermi radius $R_\mathrm{TF}=\SI{6.5}{\micro\metre}$.

In a hyperfine manifold, we expect Stark shifts to take the form~\cite{OptPolAtoms,Steck2010}
\begin{equation}
    E_{\rm S} = -\frac{1}{2} \alpha_0 \mathcal{E}^2 - \frac{1}{2} \alpha_2^F \mathcal{E}^2 \left[\frac{3 m_F^2 - F(F+1)}{F(2F-1)}\right],
\end{equation}
where $\alpha_0$ is the scalar polarizability of the ground state, $\alpha_2^F$ is the tensor polarizability for hyperfine level $F$, $m_F$ is the Zeeman sub-level, and $\mathcal{E}$ is the electric field strength. 

The scalar Stark shift will perturb the energy of both the $|F=1\rangle$ manifold and the $|F=2\rangle$ manifold equally. As a result, regardless of the position of the atoms in the trap, all atoms will simultaneously be resonant. At the centre of the cloud, both levels are shifted by about $h\times\SI{6.0}{\kilo\hertz}$. At $R_\mathrm{TF}$ from the centre, the shift is about $h\times\SI{5.9}{\kilo\hertz}$, a difference of about $h\times\SI{100}{\hertz}$.

The tensor Stark shift perturbs the energy of the $|F=1, m_F=1\rangle=|1\rangle$ state without having an effect on $|F=2\rangle$ states. As a result, atoms in the centre of the trap will have different resonant frequencies for both transitions $\omega_\A$ and $\omega_\B$ from atoms at the two ends of the trap. The magnitude of the tensor shift for \1 is $h\times\SI{25.8}{\kilo\hertz}$ at the centre of the trap and $h\times\SI{25.3}{\kilo\hertz}$ at $R_\mathrm{TF}$. The probability distribution function is~\cite{Dalfovo1999},
\begin{align}
\begin{split}
    n_\mathrm{1D}(r) &= \int_{4\pi} r^2 d\Omega\, n_\mathrm{3D}(\mathbf{r}) \\
    &= \int_{4\pi} r^2 d\Omega\frac{\mu}{g}\left(1-\frac{r^2}{R_\mathrm{TF}^2}\right) \\
    &= \frac{4\pi\mu r^2}{g}\left(1-\frac{r^2}{R_\mathrm{TF}^2}\right)
\end{split}
\end{align}
where $r$ is the radial coordinate, $m_\mathrm{Rb}$ is the mass of an \Rb{} atom, $\mu=m_\mathrm{Rb}\omega_{ho}^2R_\mathrm{TF}^2/2$ is the chemical potential, and $g=4\pi\hbar^2a/m_\mathrm{Rb}$ is the coupling constant.  We can obtain the mean Stark shift,
\begin{align}
    \langle E_{\rm S} \rangle = \int dr\, n_\mathrm{1D}(r) E_{\rm S}(r),
\end{align}
and assuming our frequency calibration is accurate to this mean, the detuning,
\begin{equation}
    \Delta(r) = E_{\rm S}(r) - \langle E_{\rm S} \rangle.
\end{equation}

To evaluate the purity and fidelity decay caused by the tensor Stark shift, we sample 1000 detunings from $\Delta(r)$, weight them by $n_\mathrm{1D}(r)$, and calculate the $\hat \rho_\mathrm{out}$ found by tomography from this distribution comparing it to the $\hat\rho$ for Stark-shift-free evolution. The results are shown in table \ref{tab:StarkShifts}.

Finally, we performed these experiments in a trap to mimic the environment of a practical quantum processor based on neutral atoms, which would require some trapping potential to maintain the atoms in position thorughout an extendend computation. However, measurements like the ones made in this work could be performed in time-of-flight (as other demonstrations have done) to avoid Stark shifts.

\begin{table}[tb!]
\caption{Simulated reduction in fidelity, purirty, and purity-adjusted fidelity due to trap-induced Stark shifts.}
\label{tab:StarkShifts}
\begin{tabular*}{3in}{@{\extracolsep{\fill}}c c c c c}
\hline
\hline
Operator & State & $\mathcal{P}$ & $\mathcal{F}$ & $\mathcal{F}_\mathrm{pure}$ \\ \hline

\multirow{3}{*}{$\hat F^{II}$} & \0 & 0.953 & 0.980 & 1.009 \\
 & \1 & 0.950 & 0.974 & 1.008 \\
 & \2 & 0.953 & 0.980 & 1.009 \\ \hline

 \multirow{3}{*}{$\hat F^{I}$} & \0 & 0.963 & 0.981 & 1.006 \\
 & \1 & 0.965 & 0.986 & 1.007 \\
 & \2 & 0.965 & 0.986 & 1.007\\
 \hline
 \hline
\end{tabular*}
\end{table}

We also perform the same purity recovery algorithm from \eqref{eq:purity-recover}, with results shown in the table. As expected, the purity recovery algorithm results in a perfect fidelity. We conclude that it is plausible that the relatively low purity and some of the error in $\mathcal{F}$ we find in the main results (table \ref{tab:rho-fidelities}) may be caused by Stark shifts from the optical dipole trap.

\subsection{Calibration}
For a high fidelity operator, the many control pulses used in our experiment including $\hat U_\mathrm{prep.}$, $\hat U_\A$, $\hat U_\B$, $\hat U_{\AB}$, and $\hat R_i$ must simultaneously have their pulse durations calibrated accurately and the detuning for both tones must be negligible (less than $0.025|\Omega|\approx 2\pi\times\SI{50}{\hertz}$ as suggested by figure \ref{fig:detuning_sim}).

To find the resonant frequencies, the duration and frequency of the two tones were adjusted manually until \SI{100}{\percent} of the atomic population was transferred between the states. The duration of each operator was also calibrated manually by starting from each computational basis state and scanning the operator duration to finding the point where the population transfer matches theory.

In the ideal case, we would assume $\Omega$ is constant during the application of an operator, and we could determine the Rabi frequency to high precision by performing long pulses $\tau=n\pi$. In our system, long pulses would not be a good calibration technique because the Rabi frequency $\Omega$ is not constant over longer time periods. This may be a limitation of our microwave amplifier or magnetic field bias stability. To account for this effect, each pulse duration was calibrated individually for its position in the pulse sequence, and only short pulses were used in the experiment.

Our current microwave system is not optimized for high Rabi frequencies. A rectangular waveguide is aimed towards the atoms from a distance of approximately 15 cm before the microwaves are allowed to propagate through free space. The microwaves which impinge on the atoms are sent off-axis and do not demonstrate any significant degree of polarization, driving $\sigma^+$, $\sigma^-$ and $\pi$ transitions. Work is underway to improve this system significantly by using a helical antenna to generate well-defined $\sigma^\pm$ polarizations based on the work of Ref. \cite{Zheng2022}.

\newpage 
\widetext
\begin{lstlisting}[
    caption={Python implementation of the decomposition algorithm.},
    morekeywords={assert, as},
    label={lst:decomp}
]
import numpy as np
from qutip import fock, rand_unitary

# Helps to check that only a diagonal Ud remains
is_diag = lambda U: not np.any((U.conj() - U.dag()).tidyup().full())


def Rotation_3D(tau, phi, basis):
    m, n = basis
    ketm = fock(3, m)
    ketn = fock(3, n)
    sig_x = ketm * ketn.dag() + ketn * ketm.dag()
    sig_y = 1j * (ketn * ketm.dag() - ketm * ketn.dag())
    return (-1j * tau / 2 * (
        np.cos(phi)*sig_x + np.sin(phi) * sig_y
        )).expm()


def decompose_step(U, basis, a):
    m, n = basis
    ketm = fock(3, m)
    ketn = fock(3, n)
    keta = fock(3, a)

    aUm = U.matrix_element(keta.dag(), ketm)
    aUn = U.matrix_element(keta.dag(), ketn)

    phi = np.pi/2 + np.angle(aUm) - np.angle(aUn)
    tau = 2 * np.arcsin(np.sqrt(
        abs(aUm) ** 2 / (abs(aUm)**2 + abs(aUn) ** 2)
        ))
    R = Rotation_3D(tau, phi, basis)
    return R, tau, phi


def decompose_UI(U):
    R1, tau1, phi1 = decompose_step(U, (0, 1), 2)
    R2, tau2, phi2 = decompose_step(U * R1.dag(), (1, 2), 2)
    R3, tau3, phi3 = decompose_step(U * R1.dag() * R2.dag(), (0, 1), 1)
    Ud = U * R1.dag() * R2.dag() * R3.dag()
    U_net = Ud * R3 * R2 * R1

    assert is_diag(Ud)
    assert not np.any((U - U_net).tidyup().full())

    return (Ud, (tau3, phi3), (tau2, phi2), (tau1, phi1))


def decompose_UII(U):
    R1, tau1, phi1 = decompose_step(U, (0, 2), 2)
    R2, tau2, phi2 = decompose_step(U * R1.dag(), (1, 2), 2)
    R3, tau3, phi3 = decompose_step(U * R1.dag() * R2.dag(), (0, 1), 1)
    Ud = U * R1.dag() * R2.dag() * R3.dag()

    U_net = Ud * R3 * R2 * R1

    assert is_diag(Ud)
    assert not np.any((U - U_net).tidyup().full())

    return (Ud, (tau3, phi3), (tau2, phi2), (tau1, phi1))


def check():
    for _ in range(10):
        U = rand_unitary(3)
        decompose_UI(U)
        decompose_UII(U)
    print('Success')


if __name__ == '__main__':
    check()
\end{lstlisting}


\end{document}